\documentclass[aps,prl,preprint,groupedaddress]{revtex4-1}
\usepackage[dvips]{graphicx}
\usepackage{amssymb,amsfonts,amsmath}
\def\be{\begin{equation}}
\def\ee{\end{equation}}
\def\bea{\begin{eqnarray}}
\def\eea{\end{eqnarray}}
\newcommand{\ft}{\tilde F}

\begin{document}

\title{The random release of phosphate controls the dynamic instability of microtubules}

\author{Ranjith Padinhateeri}
\affiliation{Department of Biosciences and Bioengineering and Wadhwani Research Centre for Biosciences and Bioengineering,
 Indian Institute of Technology Bombay, Mumbai 400076, India}

\author{Anatoly B. Kolomeisky} 
\affiliation{Department of Chemistry, Rice University, Houston, TX 77005, USA}
\author{David Lacoste}
\affiliation{ Laboratoire de Physico-Chimie Th\'{e}orique, ESPCI, 10 rue Vauquelin, Paris Cedex 05, France}
\date{\today}

\begin{abstract}
A simple stochastic model which describes microtubule dynamics and explicitly takes into account the relevant biochemical processes is presented.  The model incorporates binding and unbinding of monomers and random phosphate release inside the polymer. It is shown that this theoretical approach provides a microscopic picture of the dynamic instability phenomena of microtubules. The cap size, the concentration dependence of the catastrophe times and the delay before observing catastrophes following a dilution can be quantitatively predicted by this approach in a direct and simple way. Furthermore, the model can be solved analytically to a large extend, thus offering a valuable starting point for more refined studies of microtubules dynamics.
\end{abstract}

\maketitle

%\abbreviations{SAM, self-assembled monolayer; OTS,
%octadecyltrichlorosilane}

\noindent
\section*{Introduction}
Microtubules (MT) are involved in
key processes of cell functions such as mitosis, cell morphogenesis and motility.
The building blocks of microtubules are $\alpha \beta$-tubulin heterodimers
which can associate either laterally or longitudinally \cite{Desai_Mitchison_MT:97}.
In biological systems, microtubules display unusual non-equilibrium dynamic behaviors, which are relevant for cell functioning. One such behavior, termed treadmilling involves a flux of subunits from one polymer end to the other, and is created by a difference of critical concentrations of the two ends \cite{wilson:1998}. In another behavior, termed dynamic instability, microtubules undergo alternating phases of elongation and rapid shortening \cite{mitchison:1984}. The two behaviors, treadmilling and dynamic instability result from an interplay between the polymerization and the GTP hydrolysis.

The cap model provides a simple explanation for the dynamic instability: a growing microtubule is stabilized by a cap of unhydrolyzed units at its extremity, and when this cap is lost, the microtubule undergoes a sudden change to the shrinkage state, a so called catastrophe.
The transitions between growth and shrinking can be described
by a two state model with prescribed stochastic transitions \cite{Bayley:89,hill:84}.
This model has lead to a number of theoretical and experimental studies \cite{Verde:1992,Leibler:93,Leibler-cap:96}, which have shown in particular the existence of a phase boundary between a bounded growth and an unbounded growth regimes.
Although many features of microtubules dynamics can be captured in this way, this model remains phenomenological, because of the unknown dependence of the transition rates as function of external factors, such as tubulin concentration or temperature.

To go beyond phenomenological models, one needs to account for the main chemical
reactions occurring at the level of a single monomer \cite{margolin:2006,Wolynes:06}.
These reactions can be assumed to occur between discrete states, and the corresponding
transition rates can be observed experimentally. In this way, discrete models can be constructed,
which capture remarkably well the main dynamical features of
 single actin or single microtubule filaments \cite{kolomeisky:06,Antal-etal-PRE:07,Ranjith2009,Ranjith2010}.
These discrete models have the additional advantage of being free from some of the
limitations inherent to continuous models.

The question of the precise mechanism of hydrolysis in microtubules or actin has been
controversial for many years despite decades of experimental work.
In the vectorial model, hydrolysis occurs only at the unique interface between units bound to GTP/ATP and units bound to GDP/ADP, while
in the random model, hydrolysis can occur on any unhydrolyzed unit of the filament leading to a multiplicity of interfaces at a given time.
Between these two limits, models with an arbitrary level of cooperativity in the hydrolysis have been considered (see for instance \cite{wegner-1996,kierfeld-2010} for actin and \cite{Leibler-cap:96} for microtubules). The idea that the filament dynamics depends on the mechanism of hydrolysis in its interior or more generally on the internal
structure of the filament has been recently emphasized and it has been given the name of structural plasticity \cite{mitchison:2009}. As a practical recent illustration of that idea,  the dynamical properties of microtubules can be tuned by incorporating in them GDP-tubulin in a controlled way \cite{valiron:2010}.

In microtubules, many experimental facts point towards a
mechanism of hydrolysis which is non-vectorial but random or cooperative.
Studies of the statistics of catastrophes \cite{jason-dogterom:03,Walker-1988,voter:1991} already provided
hints about this, but there are now more direct evidences.
The observation of GTP-tubulin remnants inside a microtubule using a specific antibody \cite{perez:2008} is probably one
of the most compelling evidences.
With the development of microfluidic devices for biochemical applications,
similar experiments probing the internal structure and the dynamics of single bio-filaments are becoming more and more accessible. Furthermore, it is now possible to record the dynamics of microtubule plus-ends at nanometer resolution \cite{schek:2007,kerssemakers:2006},
thus allowing essentially to detect the addition and departure of single tubulin dimers from microtubule ends.
In view of all these recent developments,
there is a clear need to organize all this information on
microtubules dynamics with a theoretical model.
Here, we propose a simple one dimensional non-equilibrium model,
accounting for the hydrolysis occurring within the filament.
We show that this model successfully explains
for many known experimental observations with microtubules such as:
the cap size, the dependence of the catastrophe time versus monomer concentration and the delay before a catastrophe following a dilution \cite{jason-dogterom:03,Walker-1988,voter:1991}.
Our interpretation of this data confirms and goes beyond results obtained in a recent numerical and theoretical study of the dynamic instability of MT \cite{Brun-2009}.
%Finally, our model leads to predictions for the dynamics of the length following a dilution,
%which can be recorded in real time in single filament experiments.
%islands distribution of
%unhydrolyzed units within the filament, and for the size of the cap.

In vivo, the dynamics of microtubules is controlled by a variety of binding proteins,
which typically modify the polymerization process. Here we
focus on the physical principles which control the dynamic instability
of microtubules in vitro in the absence of any microtubule associated proteins.
Our model differs from previous attempts to address this problem, in that it is sufficiently simple to
be analytically solvable to a large extend, while still capturing the main features of MT dynamics.

\noindent
\section*{Model}
GTP hydrolysis is a two steps process: the first step,
the GTP cleavage produces GDP-Pi and is rapid, while the
second step, the release of the phosphate (Pi), leads to
GDP-tubulin and is by comparison much slower. This suggests that
many kinetic features of tubulin polymerization can be explained
by a simplified model of hydrolysis, which takes into account only the second
step of hydrolysis and treats tubulin subunits bound to GTP and tubulin subunits
bound to GDP-Pi as a single specie \cite{kolomeisky:06,Ranjith2009,Ranjith2010}.
This is the assumption which we make here. Therefore what we mean by random hydrolysis here
is the random process of phosphate release, which as we argue,
controls the dynamic instability of microtubules.

Our second main assumption has to do with the neglect of the protofilament structure of microtubules. Protofilaments are likely to be strongly interacting and should experience mechanical stresses in the MT lattice. We agree that modeling these effects is important to provide a complete microscopic picture of the transition from the growing phase to the shrinking phase, since this transition should involve protofilament curling near the MT ends \cite{van-buren-2005,kulic-2010}.
Here, we do not account for such effects, because as in Ref.~\cite{Leibler-cap:96}, we are interested in constructing a minimal dynamic model for microtubules, which would describe in a coarse-grained way the main aspects of the dynamics of this polymer.

\begin{figure}
\begin{center}
\includegraphics[scale=0.3]{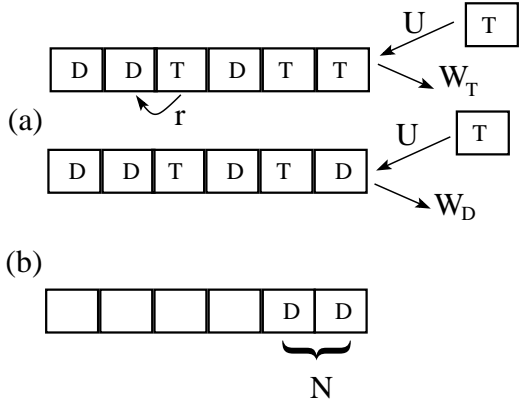}
\caption{(a): Representation of the various elementary transitions considered in the model with their corresponding rates, $U$ the on-rate of GTP-subunits, $W_T$ the off-rate of GTP-subunits, $W_D$ the off-rate of GDP-subunits and $r$ the hydrolysis rate for each unhydrolyzed unit within the filament. (b) Pattern for a catastrophe with $N$ terminal units in the GDP state.
\label{fig-sketch} }
\end{center}
\end{figure}

We also assume that the filament contains a single active end and is
in contact with a reservoir of subunits bound to GTP.
The parameters of the model are as in Refs.~\cite{kolomeisky:06,Ranjith2009,Ranjith2010}:
the rate of addition of subunits $U$,
the rate of loss of subunits bound to GTP, $W_T$, the rate of loss of
subunits bound to GDP, $W_D$, and finally the rate of GTP hydrolysis
$r$ assumed to occur randomly on any unhydrolyzed subunits within the filament.
In Fig. \ref{fig-sketch},
all these possible transitions have been depicted. We
have assumed that all the rates are independent of the
concentration of free GTP subunits $C$ except for the on-rate \cite{jason-dogterom:03},
which is $U=k_0 c$.
All the rates of this model have
been determined precisely experimentally except for $r$. The
values of these rates are given in table \ref{table-rates}.
\begin{table}
%\centering
\caption{Various rates used in the model and corresponding references. The conditions are that of a low ionic strength buffer. The value used for the rate of hydrolysis result from the analysis of the present paper. \label{table-rates}}
%\begin{tabular*}{\hsize}{@{\extracolsep{\fill}}lcrr}
\begin{tabular}{lcrr}
On-rate of T subunits at + end &$k_0$ ($\mu$M$^{-1}s^{-1}$)& 3.2 & \cite{howard-book} \cr
\hline
Off-rate of T subunits from + end & $W_T $($s^{-1}$) & 24 & \cite{jason-dogterom:03} \cr
\hline
Off-rate of D subunits from + end & $W_D$($s^{-1}$) & 290 & \cite{howard-book} \cr
\hline
Hydrolysis rate (random model) & $r$ (s$^{-1}$)& 0.2 &  \cr
\hline
%\end{tabular*}
\end{tabular}
\end{table}

As a result of the random hydrolysis, a typical filament configuration contains
many islands of unhydrolyzed subunits within the filament.
The last island containing the terminal unit is called the cap.

\noindent
\section*{Results and discussion}
%\subsection{Nucleotide content of the filament}
In this section, we obtain the nucleotide content of the filament within
 a mean-field approximation (for earlier references on this model, see \cite{Ranjith2010,kolomeisky:06,wegner:86}).
We denote by $i$ the position of a monomer within the filament, from the terminal unit at $i=1$.
For a given configuration, we introduce for each subunit
$i$ an occupation number $\tau_i$, such that $\tau_i=1$ if the subunit is bound to GTP
and $\tau_i=0$ otherwise. In the reference frame associated with the end of the filament, the equations
for the average occupation number are for $i=1$,
\be \frac{d \langle \tau_1 \rangle}{dt}=U (1- \langle \tau_1 \rangle ) - W_T \langle \tau_1 (1-\tau_2)
\rangle + W_D \langle \tau_2 (1-\tau_1) \rangle - r \langle \tau_1 \rangle, \label{recursion1} \ee
and for $i >1$,
\bea \frac{d \langle \tau_i \rangle}{dt} & = & U \langle \tau_{i-1} -\tau_i \rangle + W_T \langle \tau_1 (\tau_{i+1} - \tau_i ) \rangle \nonumber \\
& + & W_D \langle (1-\tau_1) ( \tau_{i+1} - \tau_i ) \rangle - r \langle \tau_i \rangle. \label{recursioni} \eea
In a mean-field approach, correlations are neglected, which means that for any $i,j$, $\langle \tau_i \tau_j \rangle$ is replaced by $\langle \tau_i \rangle \langle \tau_j \rangle$.
At steady state, the left-hand sides of Eqs.~\ref{recursion1}-\ref{recursioni} are both zero,
which leads to recursion relations for the $\langle \tau_i \rangle$.
Let us denote $\langle \tau_1 \rangle=q$ as the probability that the terminal
unit is bound to GTP. The recursion relations have a solution of the form for $i\geq1$,
   \be \frac{ \langle \tau_{i+1} \rangle}{\langle \tau_i \rangle}=b, \label{recursion} \ee
where $b=(U-q(W_T+r))/(U-q W_T)$.
Combining Eqs.~\ref{recursion1}-\ref{recursion}, one obtains $q$ explicitly as function of all the rates
as the solution of a cubic equation which is given in the appendix of Ref.~\cite{Ranjith2010}.
The mean filament velocity (namely the average rate of change of the total filament length) is given by
\be
v= \left( U - W_T q - W_D (1-q) \right) d,
\label{velocity}
\ee
in terms of the monomer size $d$. At the critical concentration $c_c$, the mean velocity vanishes, which corresponds to the boundary between a phase of bounded growth for $c<c_c$ and a phase of unbounded growth for $c>c_c$ \cite{Ranjith2010}.
The plot of this velocity versus concentration exhibits a kink shape near the critical
concentration, which is not particularly sensitive to the mechanism of hydrolysis since it is present both in the vectorial and random model \cite{kolomeisky:06,Ranjith2010}. This kink is well known from studies with actin \cite{hill:85} but has not been studied experimentally with microtubules except in Ref.~\cite{carlier-hill-1984} in a specific medium containing glycerol.

The distribution of the nucleotide along the filament length has a well defined steady-state in the tip reference frame at arbitrary value of the monomer concentration $c$. Using Eq.~\ref{recursion}, it follows that $\langle \tau_i \rangle= b^{i-1} q$, and therefore, the steady-state probability that the cap has exactly a length $l$, $P_l$,
is  $P_l=(\prod_{i=1}^l \langle \tau_i \rangle) (1-\langle \tau_{l+1} \rangle)$. This leads to the following expression:
\be
P_l=b^{l(l-1)/2} q^l \left( 1 - b^l q \right),
\label{SS proba}
\ee
and, the corresponding average cap size is :
\be
\langle l \rangle = \sum_{l \ge 1} l P_l = \sum_{l \ge 1} b^{l(l-1)/2} q^l.
\ee

\begin{figure}
\begin{center}
\includegraphics[scale=1]{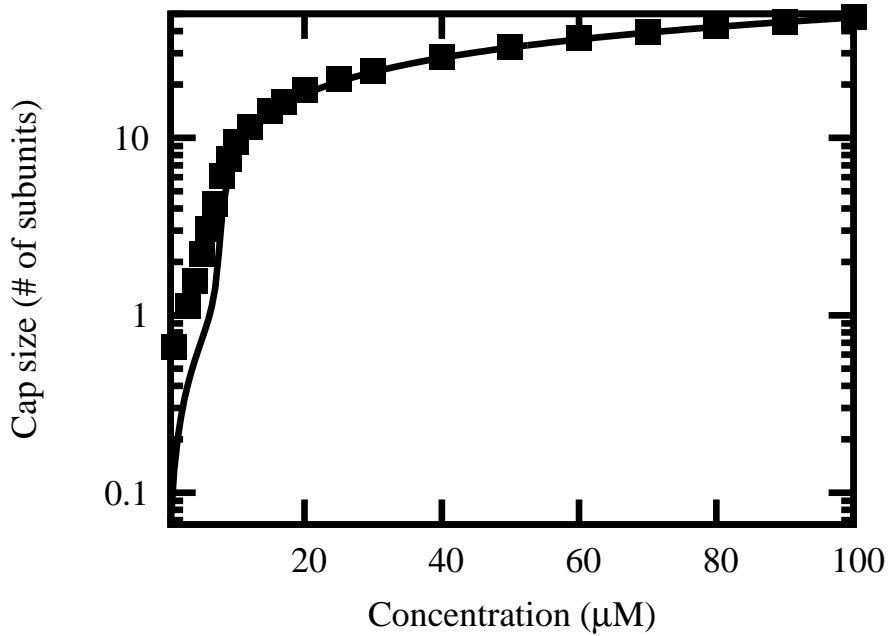}
\caption{Average cap size in number of subunits as function of the free tubulin concentration $c$ in $\mu$M. The line is the mean-field analytical solution and the
filled squares are simulation points.
\label{fig-cap} }
\end{center}
\end{figure}

In figure \ref{fig-cap}, we show how this average cap size varies as function of the free
tubulin concentration. The average cap becomes longer than approximatively one subunit above the critical concentration, $c_c$ defined above, and which is about 7$\mu$M for the parameters of table \ref{table-rates} used here. At concentrations significantly larger than this value, the cap grows more slowly, as $\sqrt{\pi U/2r}$ as $U \rightarrow \infty$ \cite{Antal-etal-PRE:07,Leibler-cap:96}. In the range of concentration [0:100 $\mu$M], the cap stays smaller than about 47 subunits, which represents only 3.6 layers (or 28 nm). This estimate indicates that the cap is below optical resolution in the range of tubulin concentration generally used, which could explain the difficulty for observing it experimentally.

A long standing view in the literature is that the cap could be as small as a single layer, as shown by experiments based on a chemical detection of the phosphate release \cite{wilson:2002}.
This view has been recently challenged by two experiments, in which the length fluctuations of microtubules were probed at the nanoscale, \cite{schek:2007,kerssemakers:2006}. The interpretation of these experiments still generate debates \cite{howard:2009,odde:2008}. In any case, taken together these two experimental studies reported a highly variable MT plus-end growth behavior, which suggests that the cap size is a fluctuating quantity, larger than one layer but smaller than about 5 layers. We note that such a range is compatible with our prediction and agrees with the estimation obtained from dilution experiments \cite{voter:1991}. Furthermore, our stochastic model naturally incorporates a fluctuating cap size. Even if the cap is indeed below optical resolution, we note that this does not rule out the possibility that it could be observed with the technique of Ref.~\cite{perez:2008}.

In figure \ref{fig-cap}, we also compare the predictions of the mean-field approximation with an exact simulation of the dynamics. We find that  mean-field theory provides an excellent approximation of the exact solution when the free tubulin concentration is above the critical concentration, which corresponds to the conditions of most experiments \cite{jason-dogterom:03,Walker-1988}. Deviations can be seen between the exact solution and its mean-field approximation in figure \ref{fig-cap} but only below the critical concentration.
Many other quantities of interest follow from the determination of the nucleotide content of a given subunit, namely $\langle \tau_i \rangle$, such as the length fluctuations of the filament \cite{Ranjith2010} or the islands distribution of hydrolyzed or non-hydrolyzed subunits \cite{Antal-etal-PRE:07,kierfeld-2010}. These predictions should prove particularly useful in testing this model against experiments, since the island distribution of unhydrolyzed units or "remnants" will become accessible in future experiments similar to that of \cite{perez:2008} but carried out in in vitro conditions.

\noindent
\subsection*{Frequency of catastrophes and rescues vs. concentration}
One difficulty in bridging the gap between a model of the dynamic instability
and experiments, lies in a proper definition of the event which is called a catastrophe, since
 the number of reported catastrophes is affected by several factors depending on the experimental conditions, such as for instance the experimental resolution of the observation \cite{schek:2007}.

Although a catastrophe manifests itself experimentally as an abrupt reduction of the total filament length, we choose to define it from the nucleotide content of the terminal region. %This approximation amounts to neglect the time associated with the loss of GDP subunits once the cap has been lost, which is typically very fast (since $W_D \gg W_T$).
Following closely Ref.~\cite{Brun-2009}, we define a shrinking configuration as one in which the last $N$ units of the filament are all in the GDP state (irrespective of the state of the other units) as shown in figure \ref{fig-sketch}.
The remaining configurations (with an unhydrolyzed cap of any size or when the number of hydrolyzed subunits at the end is less than $N$) are assumed to belong to the growing phase.
In such a two states description of the dynamics (with a growing and a shrinking phase),
which is implicitly assumed in the analysis of most experiments, the catastrophe frequency $f_c(N)$ is the inverse of the average time spent in the growing phase, while the rescue frequency $f_r(N)$ is the inverse of the average time spent in the shrinking phase. It follows from this that the catastrophe frequency $f_c(N)$ can be obtained as the probability flux out of the growing state divided by the probability to be in the growing state. For instance for $N=1$, this flux condition is
\be
\label{catastrophe frequencyN=1}
f_c(1) q = (W_T+r) P_1 + r \sum_{j \geq 2} P_j,
\ee
where the terms on the right proportional to $P_1$ correspond to a transition of the terminal unit from the GTP to the GDP state, which can occur either through hydrolysis or depolymerization of that unit, while the last term corresponds to hydrolysis of the terminal unit from cap states of length larger or equal than 2.
We have derived the general expression of $f_c(N)$ in the case of an arbitrary $N$ as shown in Supporting Information (SI) Methods, and we have checked these results by comparing them with stochastic simulations using the Gillespie algorithm \cite{gillespie:77}.

In the case of the vectorial model, the last term in Eq.~\ref{catastrophe frequencyN=1} is absent and the catastrophe frequency is non-zero only below the critical concentration. The fact that catastrophes are observed in \cite{jason-dogterom:03} significantly above the critical concentration indicates that this data is incompatible with a vectorial mechanism. For this reason, we only discuss here the predictions of the random model.

\begin{figure}
\begin{center}
\rotatebox{-90}{\includegraphics[scale=0.4]{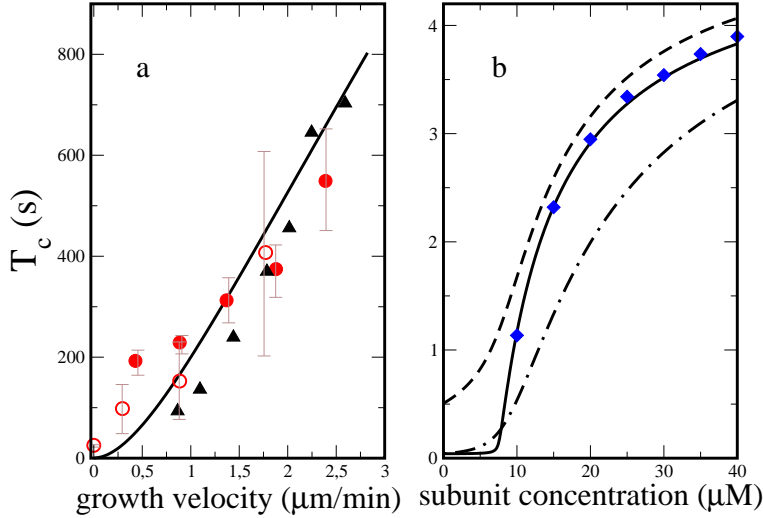}}
\caption{(a): catastrophe time $T_c$ vs growth velocity for the $N=2$ case, with the theoretical prediction (solid line) together with simulation points ($\blacktriangle$) and experimental data points taken from \cite{jason-dogterom:03} for constrained growth ($\circ$) and free growth ($\bullet$). (b): catastrophe time $T_c$ vs free subunit concentration for the $N=1$ case, with the theoretical prediction (black solid line) and simulations (filled diamonds). In addition, the dash-dotted line and the dashed line represent respectively $\langle T(1) \rangle$ and $\langle T(20) \rangle$, which have been calculated using Eq.~\ref{full Tk}.
\label{fig-N} }
\end{center}
\end{figure}

The catastrophe time $T_c(N)=1/f_c(N)$ is shown as function of growth velocity for $N=2$ in figure \ref{fig-N}a, and as a function of the concentration of free subunits, $c$, for $N=1$ in figure \ref{fig-N}b. The growth velocity is simply proportional to the concentration of free subunits. For both plots, one sees that below the critical concentration which is in the range of 5-10 $\mu$M, the catastrophe time is zero as expected since there is no stable filament in that region of concentration. Note that $T_c(N)$ behaves linearly as function of $c$ for $N=2$ but it behaves non-linearly for $N=1$. Since the experimental data of \cite{jason-dogterom:03} shows a linear dependence, this comparison indicates that the data can be explained with the model for $N=2$ but not for $N=1$. The same observation has been made in Ref.~\cite{Brun-2009}, where the same data has been analyzed. Note however, in comparing this work with this reference the following differences: first, the model of Ref.~\cite{Brun-2009} neglects rescues and assumes that the duration of a catastrophe once started is zero while the present model includes rescues, and takes into account the finite rate of loss of GDP units. Secondly the results of Ref.~\cite{Brun-2009} corresponds to the regime of high concentration of free subunits while the present model holds at any concentration even in the proximity or below the critical concentration. Thirdly the present approach leads to analytical results with the assumption that the filament has no protofilament structure while the results of Ref.~\cite{Brun-2009} are numerical but that model includes a protofilament structure.
Our analytical derivation of the catastrophe time confirms that the case $N=1$ differs in an essential way from the $N \geq 2$ case at high concentration. Indeed, the catastrophe time reaches a plateau when the concentration goes to infinity for $N=1$, while it goes to infinity for $N \geq 2$. This trend is already apparent in the figure \ref{fig-N}.

In figure \ref{fig-N}, we have used a value for the rate of hydrolysis $r=0.2$, which is higher than that estimated in Ref.~\cite{Leibler-cap:96} (there the estimate was 0.002). The reason is that the hydrolysis rate is a global factor which controls the amplitude of the catastrophe time, basically $T_c(N)$ scales for an arbitrary $N$ as $1/r^N$. The value $r=0.002$ leads to a reasonable estimate for $T_c(N)$ for $N=1$ (albeit with the wrong dependence on concentration), but if we take seriously as we do here, the observation that only the definition with $N=2$ is compatible with the measured concentration dependence of the catastrophe time, then $r$ must have a significantly larger value than expected, and
$0.2$ is the value that is needed for $T_c$ in order to match the experimental data. Finally, we also note that the scaling of $T_c(N)$ as a power law of $r$ means that large values of $N$ (such as $N>2$) can be excluded given the observed range of catastrophe times.

\begin{figure}
\begin{center}
\includegraphics[scale=1.4]{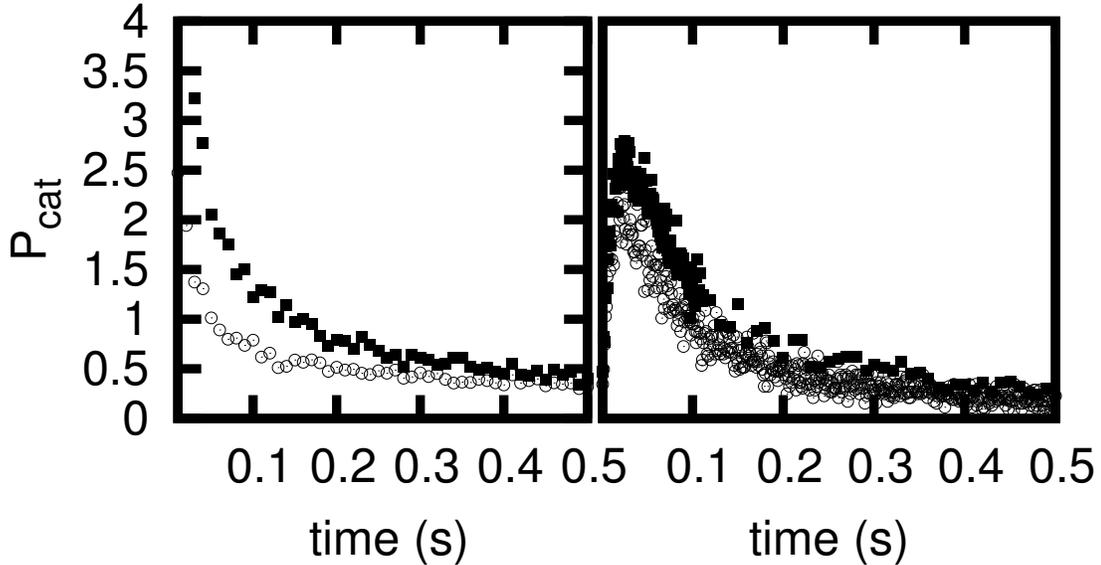}
\caption{(left)The distribution of catastrophe time (N=1) for different concentration values. $C=9\mu M$(filled squares) and $C=12\mu M$(open circles). (right) The distribution of catastrophe time (N=2) for different concentration values $C=9\mu M$(filled squares) and $C=12\mu M$(open circles). The distributions are normalized.
\label{fig:distrib-catastrophe2}}
\end{center}
\end{figure}

We also show the distribution of catastrophe times calculated with the parameters given in table \ref{table-rates}, for $N=1$ and $N=2$ in figure \ref{fig:distrib-catastrophe2}. These distributions in both cases are essentially exponential (except at a very short time which is probably inaccessible in practice in the experiments), in agreement with the observations reported in Ref.~\cite{jason-dogterom:03} with free filaments.

One advantage of our microscopic model is that it can explain and predict different related aspects of the dynamic instability of microtubules. Specifically, it also allows to predict the statistics of rescue events when the polymer switches from the shrinking phase back into the growing phase. Assuming that the system reached a steady-state behavior, the frequency of rescues $f_{r}(N)$ can be calculated using flux conditions similar to the ones used to obtain $f_c(N)$ (see SI Methods for more details). The corresponding expression is rather simple and it can be written as
\be \label{fr}
f_{r}(N)=U+W_{D} b^{N} q.
\ee
We have carried out a complete numerical test of this frequency of rescues using stochastic simulations, which is shown in SI Fig.~1.

Our model predicts that rescue events should be observable under typical cellular conditions and in experiments. However, surprisingly there is a very limited experimental information on rescues. The analysis of Eq. (\ref{fr}) might shed some light on this issue.  At low concentrations of GTP monomers in the solution, when the rate $U$ is small, the average time before the rescue event, $T_{r} \simeq 1/U$, might be very large. As a result, it might not be observable in experiments since the polymer with $L$ monomers could collapse faster ($T_{collapse} \simeq L/W_{D}$)  before any rescue event could take place. At large $U$, rescues are more frequent given that the polymer is in the shrinking state.  But the frequency of the catastrophes is very small under these conditions, the microtubule is almost always in the growing phase. Therefore in these conditions, rescues are not observed \cite{jason-dogterom:03}.

\noindent
\subsection*{First passage time of the cap and dilution experiments}
In dilution experiments, the concentration of free tubulin is abruptly reduced to a small value, resulting in catastrophes within seconds, independent of the initial concentration \cite{Walker-1991,voter:1991}. This observation is an evidence that the cap is short and independent of the initial concentration. The idea that the cap is short is also supported by the observation that cutting the end of a microtubule typically with a laser results in catastrophe. As we shall see below, all these well-known experimental facts about microtubules can be explained by the present model.

\begin{figure}
\begin{center}
\rotatebox{-90}{\includegraphics[scale=0.4]{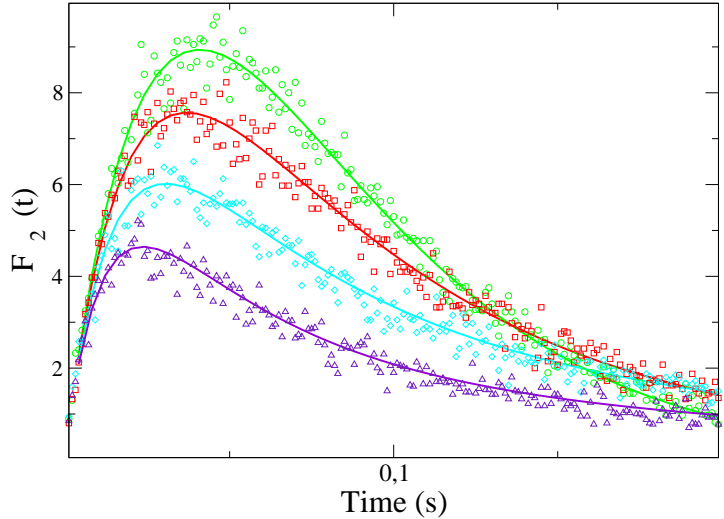}}
\caption{Distributions of the first passage time of the cap for an initial cap of $k=2$ units, $F_2(t)$ as function of the time $t$, for various initial concentration of free monomers. The solid lines are the theoretical predictions deduced from Eq.~(13) of the SI Methods after numerically inverting the Laplace transform, while the symbols are simulations. The circles correspond to a dilution into a medium with no free monomers, squares correspond similarly to a dilution into a medium with a concentration of free monomers of 2$\mu$M, diamonds to 5$\mu$M, and triangles to 9$\mu$M.
\label{fig:DFPT}}
\end{center}
\end{figure}

Here, we are interested in the time until the first catastrophe appears following the dilution. For simplicity, we take the definition of catastrophe introduced in the previous section for $N=1$, which means that a catastrophe starts as soon as the cap has disappeared (as shown in the previous section, one could extend this result to the more general case of an arbitrary $N$). Let us then introduce $F_k(t)$ the distribution of the first passage time $T_k$ for an initial condition corresponding to a cap of length $k$, and a filament in contact with a medium of arbitrary concentration.
As explained in SI Methods, it is possible to calculate analytically $F_k(t)$, by a method recently used in the context of polymer translocation \cite{PK-KironeJStatMech-2010}.
After numerically inverting the Laplace transform of $F_k(t)$,
one obtains the distribution $F_k(t)$ which is shown as solid lines in figure \ref{fig:DFPT} for the particular case of $k=2$. As can be seen in this figure, the predicted distributions agree very well with the results obtained from the stochastic simulation in this case.

\begin{figure}
\begin{center}
%\includegraphics[scale=0.3]{D1_Tres-dis}
%\caption{ The distribution of rescue time (N=1) for different concentration values. $C=9\mu M$(red),$C=12\mu M$(green) }
\rotatebox{-90}{\includegraphics[scale=0.4]{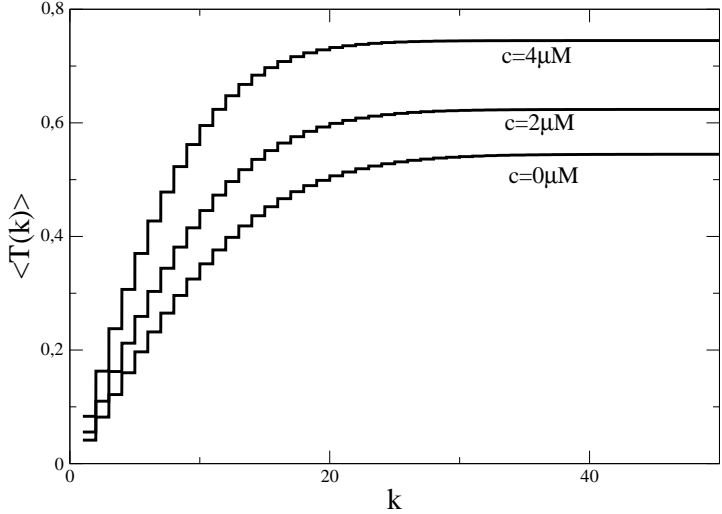}}
\caption{Mean first passage time $\langle T(k) \rangle$ as function of $k$ for three values of the monomer concentration from bottom to top 0, 2 and 4$\mu$M. The presence of steps in these curves is due to the fact that $\langle T(k) \rangle$ is only defined on integer values of $k$. Note the existence of a plateau for all values of the monomer concentration.
\label{fig:Tk}}
\end{center}
\end{figure}

From the distribution $F_k(t)$ we obtain its first moment,
the mean first passage time of the cap $\langle T(k) \rangle$. As shown in SI Methods, we find that
\be
\langle T(k) \rangle= \sum_{j=0}^{k-1} y^j \frac{J_{n+j+1}(\bar{y})}{\sqrt{U W_T} J_n(\bar{y})
- U J_{n+1}(\bar{y})},
\label{full Tk}
\ee
where $y=\sqrt{W_T/U}$, $\bar{y}=2\sqrt{U W_T}/r$, $n=(U+W_T)/r$, and the functions $J_n(y)$ are Bessel functions.
The dependance of $\langle T(k) \rangle$ as a function of the initial size of the cap $k$ is shown in figure \ref{fig:Tk}: at small $k$, $\langle T(k) \rangle$ is essentially linear in $k$ as would be expected at all $k$ in the vectorial model of hydrolysis \cite{Ranjith2009}, while here it saturates at large values of $k$ (the value of this plateau can be calculated analytically but only for $U=0$ see SI Methods). To understand this saturation, consider a cap which is initially infinitively large, then after a time of order $1/r$, the cap abruptly becomes of a finite much smaller size as a result of the hydrolysis of one unit at a random position within the filament. This feature will always happen irrespective of the monomer concentration, and indeed in figure \ref{fig:Tk}, $\langle T(k) \rangle$ has a plateau for $k \rightarrow \infty$ for all values of the monomer concentration. We note that such a behavior of
$\langle T(k) \rangle$ as function of $k$ has similarities with the case of non-compact exploration investigated in \cite{condamin:2007}, while the vectorial model of hydrolysis would correspond in the language of this reference to the case of compact exploration.

Let us now turn to a practical use of this quantity for characterizing the dynamic instability.
In the previous section, we calculated the catastrophe time $T_c$. We expect that this quantity is an average of $\langle T(k) \rangle$, and indeed we find for the case of $N=1$ that $T_c$ is bounded by $\langle T(1) \rangle$ and $\langle T(20) \rangle$ (the choice of $20$ is purely illustrative) as shown in figure \ref{fig-N}.
The characteristic time observed in dilution experiments is another average of $\langle T(k) \rangle$. More precisely, let us denote $\langle T(k) \rangle_{post}$ as the first passage time in post-dilution conditions given that the initial length of the cap is $k$. The dilution time $T_{dilution}$ is then the average of $\langle T(k) \rangle_{post}$ with respect to the steady-state probability distribution of the initial conditions before the dilution occurs. In other words,
\be
T_{dilution}=\sum_{k} \langle T(k) \rangle_{post} P_k({\rm predilution}),
\label{dilution}
\ee
where $P_k(pre-dilution)$ is the stationary probability given in Eq.~\ref{SS proba} in pre-dilution conditions.

\begin{figure}
\begin{center}
\includegraphics[scale=1]{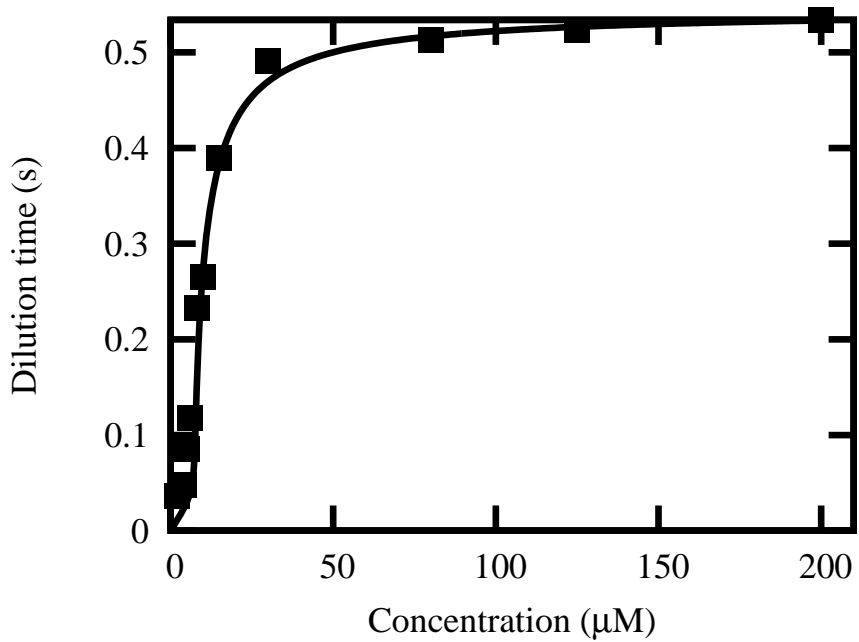}
\caption{Dilution time (s) as function of free tubulin concentration (in $\mu$M) before dilution in the case that the post-dilution tubulin concentration is zero. Solid line is the mean-field prediction based on Eq.~\ref{dilution} and the symbols are simulation points. As found experimentally, the dilution time is essentially independent of the concentration of tubulin in the pre-dilution state, and the time to observe the first catastrophe is of the order of seconds or less.
\label{fig-dilution} }
\end{center}
\end{figure}

In the case that the final medium after dilution is very dilute, one can assume that the final free tubulin concentration is zero, which allows to simplify the general expression given in Eq.~\ref{full Tk} as explained in SI Methods.
Using Eq.~\ref{dilution}, one obtains the dilution time for the parameters of the table \ref{table-rates} which is shown in figure \ref{fig-dilution}.
The figure confirms that the dilution time can be as short as a fraction of seconds in this case.
It is straightforward to extend this calculation to the case of an arbitrary value of the post-dilution medium ({\it i.e} for the case of a dilution of arbitrary strength) using the general expression derived in Eq.~\ref{full Tk}.
As the amplitude of the dilution is reduced (by increasing the post-dilution concentration), the dilution time increases as well but the general sigmoidal shape remains, with in particular a plateau at concentrations above the critical concentration. The presence of these plateaux means that the dilution time is essentially independent of the concentration of the monomers in pre-dilutions conditions as observed experimentally.
Note that the height of these plateaux scale with the hydrolysis rate. For instance, to explain the dilution times reported in \cite{Walker-1991}, one needs to use a smaller value of $r$ as given in the table because of the use of the $N=1$ definition of catastrophe. Alternatively, just as in the calculation of the catastrophe frequencies, it is possible to keep the expected large value of $r$ provided the $N=2$ definition of catastrophe is chosen. Thus, complementary information can be obtained from the catastrophe frequencies and the dilution times.

%\begin{figure}
%\begin{center}
%\includegraphics[scale=0.3]{T2_sim-mft}
%\caption{ }
%\end{center}
%\end{figure}

\noindent
\section*{Conclusion}
In this work, we have explained several important features about microtubules dynamics using a model for the random release of phosphate within the filament. The results of our mean-field approach are analytical to a large extend. With this approach we could recover some well known features of MT dynamics such as the mean catastrophe time and its distribution or the delays following a dilution, but we have also investigated much less studied aspects concerning the cap size, the role of the definition of catastrophes (via the parameter $N$) and the first passage time of the cap.
The theoretical model and ideas presented in this paper for the case of microtubules could also apply to other biofilaments such as actin or Par-M, for which the random hydrolysis model may be relevant as well. Furthermore, although the model describes a priori only single free filaments dynamics, it is also potentially useful for understanding constrained filaments,
in the broader context of force generation and force regulation by ensembles of biofilaments.
For this reason, it would be interesting to study extensions of the model to account for the various effects of MAPs on microtubules, which should shed light on the behavior of microtubules in more realistic biological conditions.
We hope that this theoretical work will stimulate further experimental and theoretical studies of these questions.

%\appendix*
\begin{acknowledgments}
We thank F. Perez, F. Nedelec and M. F. Carlier for inspiring discussions. We also would like to thank K. Mallick for pointing to us Ref.~\cite{PK-KironeJStatMech-2010}, and M. Dogterom for providing us with the data of Ref.~\cite{jason-dogterom:03}. RP acknowledges
support through IYBA, from Department of Biotechnolgy, India.
\end{acknowledgments}

~\\
\noindent
{\bf {\large Supporting Information}}
%%%%%%%%%%%%%%%%%%%%%%%%%%%%%%%%%%%%%%%%%%%%%%%%%%%%%%%%
\section{\label{AFPT} Distribution of first passage time of the cap in the random model}
%%%%%%%%%%%%%%%%%%%%%%%%%%%%%%%%%%%%%%%%%%%%%%%%%%%%%%%%
Let us denote by $F_k(t)$ the probability distribution of the first passage time of the GTP-tip (also called cap in the main text), for a cap which is initially of length $k$. This quantity obeys the following backward master equation, for $k \geq 1$,
\be
\frac{\partial F_k}{\partial t}= U (F_{k+1}- F_k ) + W_T ( F_{k-1} - F_k ) + r \left(
\sum_{j=0}^{k-1} F_j - k F_k \right).
\ee
These equations are supplemented by the boundary condition
$F_0(t)=\delta(t)$. We will assume that the random walk followed by the cap is recurrent, which means here that the disappearance of the cap is certain, whatever the time it takes. That condition means that for all $k \geq 0$,
\be
\int_0^\infty F_k(t) dt=1.
\ee
We will make use of the Laplace transform of $F_k(t)$ defined by
\be
\ft_k(s)= \int_0^\infty e^{-s t} F_k(t) dt.
\ee
With this definition, the equations above take the following form, again for $k \geq 1$
\be
(s+W_T+kr+U) \ft_k = U \ft_{k+1} + W_T \ft_{k-1} + r \sum_{j=0}^{k-1} \ft_j,
\label{general recursion}
\ee
with in addition the conditions $\ft_0(s)=1$ and for all $k \geq 0$, $\ft_k(s=0)=1$, which follows
from the normalization condition and the definition of the Laplace transform above.

For the applications of this first passage time distribution to dilution experiments, we are interested mainly in calculating it using post-dilution conditions.
In the case of a dilution, the concentration of the free monomers following dilution is in general small.
Let us discuss separately the particular case where the concentration of the medium after dilution is zero in which case $U=0$, and the general case of a dilution into a medium of prescribed concentration corresponding to $U \neq 0$.

\noindent
\subsection{Particular case of $U=0$}
In this particular case, the recursion equations given in Eq.~\ref{general recursion} are easy to solve.
The solution is
\be
\ft_k(s)=1- \frac{s}{s+W_T+r} \left( 1 + \sum_{m=1}^{k-1} \prod_{j=1}^m \frac{W_T}{s+W_T+ (j+1)r} \right),
\ee
for $k \geq 1$ with the convention that a sum over an index which ends at 0 is void.
The mean first passage time $T(k)$, is the first moment of $F_k(t)$ and thus satisfies
$\langle T(k) \rangle=-d \ft_k/ds_{s=0}$. It follows that for $k \geq 1$,
\be
\langle T(k) \rangle=\frac{1}{W_T +r } \left( 1 + \sum_{m=1}^{k-1} \prod_{j=1}^m \frac{W_T}{W_T+ (j+1)r} \right).
\label{Tk}
\ee

In this particular case of $U=0$, it is possible to derive an asymptotic form of this mean first passage time for $k \rightarrow \infty$, namely $\langle T \rangle=\lim_{k \rightarrow \infty} \langle T(k) \rangle$ . Indeed in this case, the sum can be written in terms of hypergeometric functions \cite{Antal-etal-PRE:07,abramowitz}, and it reads
\be
\langle T \rangle = \frac{1}{W_T+r} F(1;\frac{W_T}{r}+2;\frac{W_T}{r}).
\ee

The expression of the mean-first passage time given in Eq.~\ref{Tk} can be used
to obtain the delay before the appearance of the first catastrophe as explained in the
main text.
In this case, we find
\be
T_{dilution}=\sum_{k \geq 1} \Delta_k b^{k(k-1)/2} q^k,
\ee
where
\be
\Delta_k=\frac{1}{W_T + r} \prod_{j=1}^{k-1} \frac{W_T}{W_T + (j+1) r}.
\ee
This dilution time is shown in Fig.~7 of the main text.

\noindent
\subsection{General case of $U \neq 0$}
The solution to this general case is more involved but it can be obtained using Bessel functions (for a solution of a similar recursion see \cite{PK-KironeJStatMech-2010}). In a first step, we transform the recursion of Eq.~\ref{general recursion} using the difference variable $K_k(s)=\ft_k(s) - \ft_{k+1}(s)$, which leads to
\be
(s+W_T+(k+1)r+U) K_k(s) = U K_{k+1}(s) + W_T K_{k-1}(s).
\label{difference eq}
\ee
Then, we introduce the change of variable $K_k(s)=y^k g_k(s)$ and we choose $y=\sqrt{W_T/U}$ in such a way that Eq.~\ref{difference eq} takes the simpler form:
\be
g_{k+1}(s)+g_{k-1}(s)=\frac{s+U+W_T+(k+1)r}{\sqrt{U W_T}} g_k(s).
\ee
The solution to this equation can be obtained by comparing with the well-known identity
\be
J_{\nu+1}(x)+J_{\nu-1}(x) = \frac{2 \nu}{x} J_\nu (x),
\ee
for Bessel functions.
Thus, the solution has the form
\be
g_k(s)=C J_{(s+U+W_T+(k+1)r)/r}(\bar{y}),
\ee
where $C$ is a constant and $\bar{y}=2 \sqrt{U W_T}/r$. The boundary condition given above for $\ft_0(s)$
leads to the following condition
\be
g_0(s)=\frac{U y g_1(s) + s}{s+W_T+r},
\ee
which fixes the constant $C$. In the end, one obtains
\be
g_k(s)=\frac{s J_{(s+U+W_T+(k+1)r)/r}(\bar{y})}{-U J_{(s+U+W_T+r)/r}(\bar{y})+ \sqrt{U W_T} J_{(s+U+W_T)/r}(\bar{y})},
\ee
which satisfies in addition the required condition at $s=0$ namely that for all $k \ge 0$, $g_k(s=0)=0$. With this expression, one obtains
the Laplace transform of the first passage distribution of the cap, $\ft_k(s)$ from
\be
\ft_k(s)=1- \sum_{j=0}^{k-1} y^j g_j(s).
\label{general Fk}
\ee
Although it is not immediately apparent, it can be checked that the particular case discussed above is indeed recovered by taking the limit $U \rightarrow 0$ of the general case.
After using $\langle T(k) \rangle=-d \ft_k/ds_{s=0}$ together with Eq.~\ref{general Fk}, one obtains the general expression for the mean first passage time of the cap $\langle T(k) \rangle$
given in the main text, which reads
\be
\langle T(k) \rangle= \sum_{j=0}^{k-1} y^j \frac{J_{n+j+1}(\bar{y})}{\sqrt{U W_T} J_n(\bar{y})
- U J_{n+1}(\bar{y})},
\label{full Tk}
\ee
where $n=(U+W_T)/r$.

%%%%%%%%%%%%%%%%%%%%%%%%%%%%%%%%%%%%%%%%%%%%%%%%%%%%%%%%
\noindent
\section{\label{catastrophe} Catastrophes and rescues for arbitrary $N$}
%%%%%%%%%%%%%%%%%%%%%%%%%%%%%%%%%%%%%%%%%%%%%%%%%%%%%%%%
Catastrophes are associated with stochastic transitions between growing and shrinking dynamic phases. The microtubule is in the growing phase when it is found in one of polymer configurations with the unhydrolyzed cap of any size or when the number of already hydrolyzed monomers at the end is less than  $N$. We define  $R_{k,l}$ as a probability to be in the polymer configuration with $l$ T monomers at the end that are preceded by $k$ D monomers (irrespective of the state of the other subunits), $Q_{k,l}$ as a probability to be in the polymer configuration with $l$ D monomers at the end that are preceded by $k$ T monomers, and finally $S_{k,l}$ is a probability that the last $l$ monomers at the end are hydrolyzed except for the one subunit at position $k$ counting from the end of the polymer. Formally these definitions can be also written as,
\bea \label{probabilities}
 R_{k,l} \equiv Prob(\ldots \underbrace{D \ldots D}_{k},\underbrace{T \ldots T}_{l}), & Q_{k,l} \equiv Prob(\ldots \underbrace{T \ldots T}_{k},\underbrace{D \ldots D}_{l}), & \nonumber \\
 S_{k,l} \equiv Prob(\ldots \underbrace{D \ldots T \ldots D}_{l}). & &
\eea
Note that the probability $P_{l}$ to have the unhydrolyzed cap of exactly $l$ monomers can be expressed as $P_{l}=R_{1,l}$, while the probability to be found in the growing phase is
\be \label{Pgr}
P_{gr}=\sum_{l=1}^{\infty} R_{1,l}+ \sum_{l=1}^{N-1} Q_{1,l}.
\ee

The simple mean-field theory assumes that the state of the monomer in the microtubule is independent of its neighbors, and it also estimates that the probability to find T or D monomer $k$ sites away from the polymer is equal to $b^{k-1}q$ or $(1- b^{k-1}q)$ respectively, with the parameter $b$ given by
\be
b=\frac{U-q(W_{T}+r)}{U-q W_{T}}.
\ee
The probabilities defined in Eq. (\ref{probabilities}) can be easily calculated yielding,
\bea \label{RQS}
R_{k,l}= b^{l(l-1)/2} q^{l} \prod_{j=l}^{l+k-1} (1-b^{j}q),  Q_{k,l}=  b^{k(2l+k-1)/2} q^{k} \prod_{j=1}^{l} (1-b^{j-1}q), &  & \nonumber \\
  S_{k,l}= b^{k-1} q \prod_{j=1}^{k-1} (1-b^{j-1}q) \prod_{j=k+1}^{l} (1-b^{j-1}q). & &
\eea
Then the probability to be found in the growing phase is
\be \label{Pgr1}
P_{gr}=q+\sum_{k=1}^{N-1} b^{k}q \prod_{j=1}^{k}(1-b^{j-1}q).
\ee

The frequency of catastrophes $f_{c}(N)$ in steady-state conditions can be found from the fact that the total flux out of the growing phase, $f_{c} P_{gr}$, must be equal to the flux to the shrinking phase, leading to the following equation,
\be
f_{c} (N) P_{gr}= W_{T} R_{N,1} + r \sum_{k=1}^{N} S_{k,N}.
\ee
Using Eqs. (\ref{RQS}) and (\ref{Pgr1}), it can be shown that
\be \label{fc}
f_{c}(N)=\frac{W_{T} \prod_{j=1}^{N}(1-b^{j}q)+r\sum_{k=1}^{N} b^{k-1}\prod_{j=k+1}^{N}(1-b^{j-1}q) \prod_{j=1}^{k-1}(1-b^{j-1}q)}{1+ \sum_{k=1}^{N-1} b^{k}\prod_{j=1}^{k}(1-b^{j-1}q)}.
\ee
For $N=1$, we obtain a simple expression for the frequency of catastrophes,
\be
f_{c}(1)=W_{T}(1-b q) +r,
\ee
while for $N=2$ it gives
\be
f_{c}(2)=\frac{W_{T}(1-bq)(1-b^{2}q) +r\left[1-bq +b(1-q) \right]}{1+b(1-q)}.
\ee
A limiting behavior of the frequency of catastrophes for general $N$ can be analyzed. For low concentrations of free GTP monomers in the solution, corresponding to $u \rightarrow 0$, we have $q \rightarrow 0$ and $b \rightarrow 1+r/w_{T}$, producing
\be
f_{c}(N) \simeq r+ \frac{W_{T}}{1+\sum_{k=1}^{N-1} (1+r/W_{T})^{k}}.
\ee
For large $N$ and small hydrolysis rates ($r/W_{T} \ll 1$) the expression for the frequency of catastrophes is even simpler,
 \be
f_{c}(N) \simeq r+ \frac{W_{T}}{N}.
\ee
Another limit of interest corresponds to large concentrations ($U \gg 1$), where $q \rightarrow 1$ and $b \rightarrow 1$, leading to $ f_{c} (N) \rightarrow 0$ for all values of $N \ge 2$, while for $N=1$ we have $f_{c} (1) \rightarrow  r$.

This method of analyzing catastrophes can be also extended to calculating frequency of rescue events $f_r(N)$. The probability to find the microtubule in the shrinking phase is equal to
\be
P_{sh}=1-P_{gr}= 1-q-\sum_{k=1}^{N-1} b^{k}q \prod_{j=1}^{k}(1-b^{j-1}q).
\ee
The total flux out of this state is given by
\be
f_{r}(N) P_{sh}= U P_{sh} +W_{D} Q_{1,N},
\ee
which leads to the following equation
\be
f_{r}(N)=U+\frac{W_{D}b^{N}q \prod_{j=1}^{N}(1-b^{j-1}q)}{1-q-\sum_{k=1}^{N-1} b^{k}q \prod_{j=1}^{k}(1-b^{j-1}q)}.
\ee
This expression can be further simplified to obtain the final result,
\be
f_{r}(N)= U+ W_{D} b^{N} q.
\label{rescue freq}
\ee
For all values of $N$ in the limit of $U  \rightarrow 0$ it yields $f_{r} \simeq U$, while for large $U$ we have $f_{r} \simeq U+W_{D}$.

In addition, the average time before the catastrophe or before the rescue can be easily obtained by inverting the corresponding expressions for frequencies, namely, $T_{c}(N) = 1/f_{c}(N)$ and $T_{r}(N) = 1/f_{r}(N)$.

\noindent
\subsection{Numerical test of the predictions for rescues frequency}
\begin{figure}
\begin{center}
\includegraphics[scale=1]{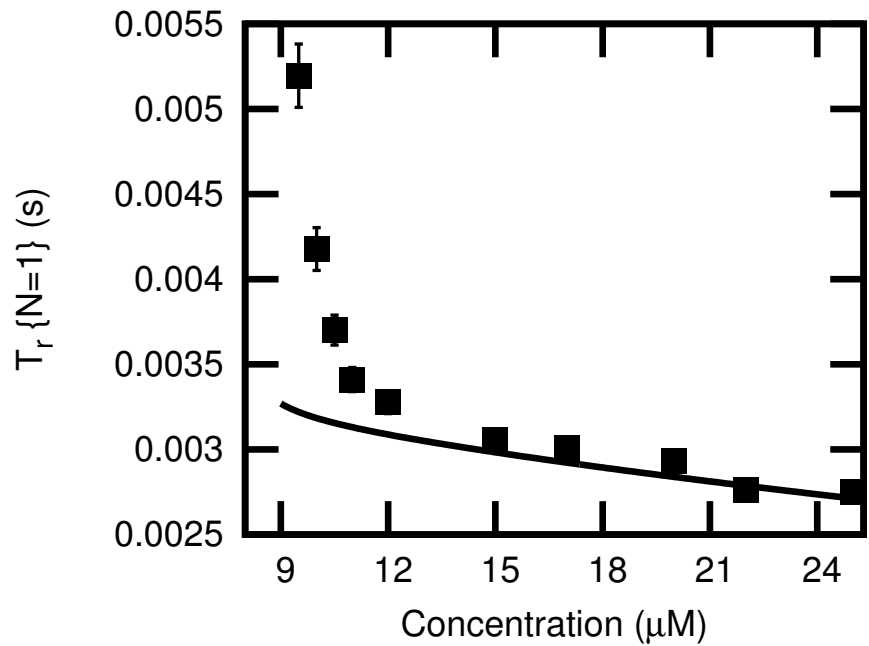}
\caption{Rescue time (N=1) as function of concentration. The solid line is obtained from the mean-field theory. The data points (filled squares) are obtained from the simulations. \label{fig:rescue-time}}
\end{center}
\end{figure}

We show here a comparison between the theoretical mean-field prediction for 
the rescue frequency given by Eq.~\ref{rescue freq} and results from stochastic simulations
in figure \ref{fig:rescue-time}.
In the conditions of this figure, the filaments are sufficiently long and thus they do not collapse before rescue events occur. The theoretical mean-field predictions agree well with the simulations at concentrations of free monomers larger than the critical concentration. Deviations are observed at low concentrations near the critical concentration, in a way which has similarities with the deviations observed in Fig.~2.
\vspace{2cm}
%\bibliography{ref_actin}

\begin{thebibliography}{0}%
\makeatletter
\providecommand \@ifxundefined [1]{%
 \@ifx{#1\undefined}
}%
\providecommand \@ifnum [1]{%
 \ifnum #1\expandafter \@firstoftwo
 \else \expandafter \@secondoftwo
 \fi
}%
\providecommand \@ifx [1]{%
 \ifx #1\expandafter \@firstoftwo
 \else \expandafter \@secondoftwo
 \fi
}%
\providecommand \natexlab [1]{#1}%
\providecommand \enquote  [1]{``#1''}%
\providecommand \bibnamefont  [1]{#1}%
\providecommand \bibfnamefont [1]{#1}%
\providecommand \citenamefont [1]{#1}%
\providecommand \href@noop [0]{\@secondoftwo}%
\providecommand \href [0]{\begingroup \@sanitize@url \@href}%
\providecommand \@href[1]{\@@startlink{#1}\@@href}%
\providecommand \@@href[1]{\endgroup#1\@@endlink}%
\providecommand \@sanitize@url [0]{\catcode `\\12\catcode `\$12\catcode
  `\&12\catcode `\#12\catcode `\^12\catcode `\_12\catcode `\%12\relax}%
\providecommand \@@startlink[1]{}%
\providecommand \@@endlink[0]{}%
\providecommand \url  [0]{\begingroup\@sanitize@url \@url }%
\providecommand \@url [1]{\endgroup\@href {#1}{\urlprefix }}%
\providecommand \urlprefix  [0]{URL }%
\providecommand \Eprint [0]{\href }%
\providecommand \doibase [0]{http://dx.doi.org/}%
\providecommand \selectlanguage [0]{\@gobble}%
\providecommand \bibinfo  [0]{\@secondoftwo}%
\providecommand \bibfield  [0]{\@secondoftwo}%
\providecommand \translation [1]{[#1]}%
\providecommand \BibitemOpen [0]{}%
\providecommand \bibitemStop [0]{}%
\providecommand \bibitemNoStop [0]{.\EOS\space}%
\providecommand \EOS [0]{\spacefactor3000\relax}%
\providecommand \BibitemShut  [1]{\csname bibitem#1\endcsname}%
\let\auto@bib@innerbib\@empty
%</preamble>
\end{thebibliography}%


\begin{thebibliography}{100}
\bibitem{Desai_Mitchison_MT:97}
Desai A, Mitchison TJ
\newblock (1997) Microtubule polymerization dynamics.
\newblock \emph{Annual Review of Cell and Developmental Biology} 13:83--117.

\bibitem{wilson:1998}
Margolis RL, Wilson L
\newblock (1998) Microtubule treadmilling: what goes around comes around.
\newblock \emph{BioEssays} 20:830.

\bibitem{mitchison:1984}
Mitchison T, Kirschner M
\newblock (1984) Dynamic instability of microtubule growth.
\newblock \emph{Nature} 312:237--242.

\bibitem{Bayley:89}
Bayley P, Schilstra M, Martin S
\newblock (1989) {A simple formulation of microtubule dynamics: quantitative
  implications of the dynamic instability of microtubule populations in vivo
  and in vitro}.
\newblock \emph{J Cell Sci} 93:241--254.

\bibitem{hill:84}
Hill TL
\newblock (1984) {Introductory analysis of the GTP-cap phase-change kinetics at
  the end of a microtubule}.
\newblock \emph{Proc. Natl. Acad. Sci. USA.} 81:6728--32.

\bibitem{Verde:1992}
Verde F, Dogterom M, Stelzer E, Karsenti E, Leibler S
\newblock (1992) Control of microtubule dynamics and length by cyclin a- and
  cyclin b-dependent kinases in xenopus egg extracts.
\newblock \emph{The Journal of Cell Biology} 118:1097--1108.

\bibitem{Leibler:93}
Dogterom M, Leibler S
\newblock (1993) Physical aspects of the growth and regulation of microtubule
  structures.
\newblock \emph{Phys. Rev. Lett.} 70:1347--1350.

\bibitem{Leibler-cap:96}
Flyvbjerg H, Holy TE, Leibler S
\newblock (1996) Microtubule dynamics: Caps, catastrophes, and coupled
  hydrolysis.
\newblock \emph{Phys. Rev. E} 54:5538--5560.

\bibitem{margolin:2006}
Margolin G, Gregoretti IV, Goodson HV, Alber MS
\newblock (2006) Analysis of a mesoscopic stochastic model of microtubule
  dynamic instability.
\newblock \emph{Phys. Rev. E} 74:041920.

\bibitem{Wolynes:06}
Zong C, Lu T, Shen T, Wolynes PG
\newblock (2006) Nonequilibrium self-assembly of linear fibers: microscopic
  treatment of growth, decay, catastrophe and rescue.
\newblock \emph{Physical Biology} 3:83--92.

\bibitem{kolomeisky:06}
Stukalin EB, Kolomeisky AB
\newblock (2006) {ATP Hydrolysis Stimulates Large Length Fluctuations in Single
  Actin Filaments}.
\newblock \emph{Biophys. J.} 90:2673--2685.

\bibitem{Antal-etal-PRE:07}
Antal T, Krapivsky PL, Redner S, Mailman M, Chakraborty B
\newblock (2007) Dynamics of an idealized model of microtubule growth and
  catastrophe.
\newblock \emph{Phys. Rev. E.} 76:041907.

\bibitem{Ranjith2009}
Ranjith P, Lacoste D, Mallick K, Joanny JF
\newblock (2009) {Nonequilibrium Self-Assembly of a Filament Coupled to ATP/GTP
  Hydrolysis}.
\newblock \emph{Biophys. J.} 96:2146--2159.

\bibitem{Ranjith2010}
Ranjith P, Mallick K, Joanny JF, Lacoste D
\newblock (2010) {Role of ATP hydrolysis in the Dynamics of a single actin
  filament}.
\newblock \emph{Biophys. J.} 98:1418--1427.

\bibitem{wegner-1996}
Pieper U, Wegner A
\newblock (1996) {The end of a polymerizing actin filament contains numerous
  ATP-subunit segments that are Disconnected by ADP-subunits resulting from ATP
  hydrolysis}.
\newblock \emph{Biochemistry} 35:4396.

\bibitem{kierfeld-2010}
Li X, Lipowsky R, Kierfeld J
\newblock (2010) Coupling of actin hydrolysis and polymerization: Reduced
  description with two nucleotide states.
\newblock \emph{Europhys. Lett.} 89:38010.

\bibitem{mitchison:2009}
Kueh HY, Mitchison TJ
\newblock (2009) {Structural Plasticity in Actin and Tubulin Polymer Dynamics}.
\newblock \emph{Science} 325:960--963.

\bibitem{valiron:2010}
Valiron O, Arnal I, Caudron N, Job D
\newblock (2010) {GDP}-{T}ubulin incorporation into growing microtubules
  modulates polymer stability.
\newblock \emph{J. Biol. Chem.} 285:17507.

\bibitem{jason-dogterom:03}
Janson ME, de~Dood ME, Dogterom M
\newblock (2003) {Dynamic instability of microtubules is regulated by force}.
\newblock \emph{J. Cell Biol.} 161:1029--1034.

\bibitem{Walker-1988}
Walker RA, {et~al.}
\newblock (1988) {Dynamic instability of individual microtubules analyzed by
  video light microscopy: rate constants and transition frequencies.}
\newblock \emph{J Cell Biol} 107:1437--1448.

\bibitem{voter:1991}
Voter W, O'Brien E, Erickson H
\newblock (1991) Dilution-induced disassembly of microtubules: relation to
  dynamic instability and the {GTP} cap.
\newblock \emph{Cell Motil Cytoskeleton.} 18:55.

\bibitem{perez:2008}
Dimitrov A, {et~al.}
\newblock (2008) {Detection of GTP-Tubulin Conformation in Vivo Reveals a Role
  for GTP Remnants in Microtubule Rescues}.
\newblock \emph{Science} 322:1353--1356.

\bibitem{schek:2007}
Schek HT, Gardner MK, Cheng J, Odde DJ, Hunt AJ
\newblock (2007) Microtubule assembly dynamics at the nanoscale.
\newblock \emph{Curr. Biol.} 17:1445.

\bibitem{kerssemakers:2006}
Kerssemakers JWJ, {et~al.}
\newblock (2006) Assembly dynamics of microtubules at molecular resolution.
\newblock \emph{Nature} 442:709.

\bibitem{Brun-2009}
Brun L, Rupp B, Ward JJ, N\'{e}d\'{e}lec F
\newblock (2009) {A theory of microtubule catastrophes and their regulation}.
\newblock \emph{Proc Natl Acad Sci USA} 106:21173--21178.

\bibitem{van-buren-2005}
Van~Buren V, Cassimeris L, Odde D
\newblock (2005) {Mechanochemical model of microtubule structure and
  self-assembly kinetics}.
\newblock \emph{Biophys J} 89:2911--2926.

\bibitem{kulic-2010}
Mohrbach H, Johner A, Kuli\ifmmode~\acute{c}\else \'{c}\fi{} IM
\newblock (2010) Tubulin bistability and polymorphic dynamics of microtubules.
\newblock \emph{Phys. Rev. Lett.} 105:268102.

\bibitem{wegner:86}
Keiser T, Schiller A, Wegner A
\newblock (1986) Nonlinear increase of elongation rate of actin filaments with
  actin monomer concentration.
\newblock \emph{Biochemi} 25:4899--4906.

\bibitem{hill:85}
Pantaloni D, Hill TL, Carlier MF, Korn ED
\newblock (1985) {A model for actin polymerization and the kinetic effects of
  ATP hydrolysis}.
\newblock \emph{Proc. Natl. Acad. Sci. USA.} 82:7207--7211.

\bibitem{carlier-hill-1984}
Carlier MF, Hill TL, Chen Y
\newblock (1984) Interference of {GTP} hydrolysis in the mechanism of
  microtubule assembly: An experimental study.
\newblock \emph{Proc. Natl. Acad. Sci. USA} 81:771.

\bibitem{wilson:2002}
Panda D, Miller HP, Wilson L
\newblock (2002) Determination of the size and chemical nature of the
  stabilizing cap at microtubule ends using modulators of polymerization
  dynamics.
\newblock \emph{Biochemistry} 41:1609--1617
\newblock PMID: 11814355.

\bibitem{howard:2009}
Howard J, Hyman AA
\newblock (2009) Growth, fluctuation and switching at microtubule plus ends.
\newblock \emph{Nature reviews} 10:569.

\bibitem{odde:2008}
Gardner MK, Hunt AJ, Goodson HV, Odde DJ
\newblock (2008) Microtubule assembly dynamics: New insights at the nanoscale.
\newblock \emph{Curr. Opin. Cell. Biol.} 20:64.

\bibitem{gillespie:77}
Gillespie DT
\newblock (1977) Exact stochastic simulation of coupled chemical reactions.
\newblock \emph{J. Phys. Chem.} 81:2340.

\bibitem{Walker-1991}
Walker RA, Pryer NK, Salmon ED
\newblock (1991) {Dilution of individual microtubules observed in real time in
  vitro: evidence that cap size is small and independent of elongation rate.}
\newblock \emph{The Journal of Cell Biology} 114:73--81.

\bibitem{PK-KironeJStatMech-2010}
Krapivsky PL, Mallick K
\newblock (2010) Fluctuations in polymer translocation.
\newblock \emph{Journal of Statistical Mechanics: Theory and Experiment}
  2010:P07007.

\bibitem{condamin:2007}
Condamin S, Benichou O, Tejedor V, Voituriez R, Klafter J
\newblock (2007) First-passage times in complex scale-invariant media.
\newblock \emph{Nature} 450:06201.

\bibitem{howard-book}
Howard J
\newblock (2001) \emph{Mechanics of Motor Proteins and the Cytoskeleton}
\newblock (Sinauer Associates, Inc., Massachusetts).

\end{thebibliography}

\begin{thebibliography}{1}

\bibitem{Antal-etal-PRE:07}
Antal T, Krapivsky PL, Redner S, Mailman M, Chakraborty B
\newblock (2007) Dynamics of an idealized model of microtubule growth and
  catastrophe.
\newblock \emph{Phys. Rev. E.} 76:041907.

\bibitem{abramowitz}
Abramamowitz M, Stegun IA
\newblock (1972) \emph{{Handbook of Mathematical Functions}}
\newblock (Dover, New York).

\bibitem{PK-KironeJStatMech-2010}
Krapivsky PL, Mallick K
\newblock (2010) Fluctuations in polymer translocation.
\newblock \emph{Journal of Statistical Mechanics: Theory and Experiment}
  2010:P07007.

\end{thebibliography}

\end{document}